\documentclass[preprint,12pt]{elsarticle}
\usepackage{lineno,hyperref}
\modulolinenumbers[5]
\usepackage{amssymb}
\usepackage{graphicx}
\usepackage{tikz}
\usepackage{amsmath}
\journal{ }
\usepackage{pdf14}

\begin{document}

\begin{frontmatter}

\title{Characterization of Hybrid Modes\\ in Metamaterial Waveguides}

\author{M.~Beig-Mohammadi$^{1}$, Nafiseh Sang-Nourpour$^{1, 2, 3}$, Barry C.~Sanders$^{2, 4, 5, 6}$, Benjamin R.~Lavoie$^{7, 2}$, R.~Kheradmand$^{1,*}$}

\address{$^{1}$Photonics group, Research Institute for Applied Physics and Astronomy, University of Tabriz, Iran}
\address{$^{2}$Institute for Quantum Science and Technology, University of Calgary, Alberta, Canada}
\address{$^{3}$Photonics Group, Aras International Campus of University of Tabriz, Iran}
\address{$^{4}$Hefei National Laboratory for Physical Sciences at the Microscale,
University of Science and Technology of China, Hefei, Anhui, China}
\address{$^{5}$Shanghai Branch, CAS Center for Excellence and Synergetic Innovation Center in Quantum Information and Quantum Physics, University of Science and Technology of China, Shanghai, China}
\address{$^{6}$Program in Quantum Information Science, Canadian Institute for Advanced Research, Toronto, Ontario, Canada}
\address{$^{7}$Department of Electrical and Computer Engineering, University of Calgary, Alberta, Canada}
\ead{$^{*}$r\underline{ }kheradmand@tabrizu.ac.ir}

\begin{abstract}
In this paper, we employ the properties of metamaterials to tailor the modes of metamaterial-dielectric waveguides operating at optical frequencies.
We survey the effect of fishnet metamaterial structural parameters such as the magnetic oscillation strength, magnetic resonance frequency and magnetic damping
on the double-negative refractive index frequency region in metamaterials and on the hybrid-modes in slab metamaterial-dielectric waveguides.
To identify the robustness of the metamaterials
to fluctuations in the metamaterial structural parameters,
we investigate the behavior of metamaterials under Gaussian errors on their structural parameters.
Our survey enables the identification of appropriate metamaterial unit-cell structure and the permissive fluctuations on the structural parameters for further applications of metamaterials in waveguide technologies.
\end{abstract}


\end{frontmatter}

\section{Introduction}

Metamaterials are artificially engineered materials
with properties such as negative refractive index \cite{veselago2003electrodynamics} and
perfect absorption \cite{landy2008perfect}.
Metamaterials have promising applications
ranging from perfect lenses \cite{pendry2000negative,chen2012metamaterials} and cloaking devices \cite{cai2007optical,pendry2006controlling} to electromagnetic (EM) waveguides \cite{yeh2008essence,sang2015electromagnetic}.
EM waveguides are able to confine and direct the energy of EM fields \cite{tong2014advanced} 
with applications in radar and optical fiber \cite{luo2002all}.
In recent years,
waveguide properties are improved
by engineering metamaterials in the waveguide structures \cite{sang2015electromagnetic,d2005te,wang2007nanoscale}.
Different propagation characteristics and new kinds of modes \cite{sang2015electromagnetic,lavoie2012low} 
make metamaterial waveguides appropriate candidates for guiding EM waves.
One advantage of metamaterials is that their electromagnetic properties are tailorable to suit the desired application.
Tailoring capabilities of metamaterials lies on varying the metamaterial structural parameters.

Metamaterial tailoring methods can be divided into three classes:
(i)~circuit tailoring methods to control the resonance frequency of the unit-cells \cite{shadrivov2012metamaterials,slobozhanyuk2014nonlinear,rose2011overcoming,lapine2004three, powell2007self};
(ii)~geometrical tailoring methods,
which allow the modification of EM response by changing the geometry
(size, orientation, period, etc.) of the unit-cell \cite{lapine2009structural,liu2012micromachined}; 
(iii)~material tailoring methods
where EM response of the material is modified under a suitable external stimulus
like light, electric and magnetic field \cite{rizza2015reconfigurable}. 

We theoretically address the first two methods of tailoring metamaterial properties, namely circuit and geometrical tailoring.
As the propagation characteristics of waveguides
are affected by the properties of materials in the core and cladding \cite{tong2014advanced},
the proper choice of waveguide material parameters is important.
For the fishnet unit-cell metamaterials considered in this paper,
the effective structural parameters are
magnetic resonance frequency, magnetic damping and magnetic oscillation strength \cite{penciu2010magnetic}.
Fishnet metamaterials allow for a higher resonance frequency,
compared to other known metamaterial unit-cells,
which makes these metamaterials appropriate candidates at optical frequencies \cite{penciu2010magnetic}.

Specifically,
we theoretically investigate the effects of metamaterial structural parameters
on the negative refractive index frequency region of metamaterials
and analyze the robustness of the metamaterials to the
inaccuracy of their structural parameters.
We also investigate the effect of structural parameters
on the modal behavior of slab metamaterial-dielectric waveguide
with metamaterial cladding and dielectric core.
By tailoring the metamaterial EM response,
we can control modal characteristics of metamaterial waveguides
according to the requirements.
Our survey helps in choosing
appropriate parameters when constructing these kinds of waveguides. 

To characterize the EM responses of metamaterials to incident field,
we investigate the behavior of refractive index.
We analyze the modal behavior of metamaterial waveguides
by employing metamaterials with different structural parameters.
The relevant background
of our approach is presented in Sec.~2.~We detail our results
in Sec.~3 and we conclude with a discussion
of our theory in Sec.~4.

\section{Theory and Approach}

This section provides the relevant background and approach required
to survey waveguide behavior
considering metamaterial structural parameters.
We begin with
a brief review of EM susceptibilities of fishnet metamaterials
and discuss the effective magnetic parameters in tailoring EM responses of metamaterial and the robustness of metamaterials to the inaccuracy of their structural parameters.
We then present
the characteristics of slab waveguide
and discuss the modal behavior of metamaterial-dielectric slab waveguides.

\subsection{Electromagnetic Susceptibilities}

The permittivity of metamaterials is described by the Drude model~\cite{cai2010optical}:
\begin{equation}
\begin{split}
\frac{\varepsilon(\omega)}{\varepsilon_{0}}
=\varepsilon'(\omega)+\text{i} \varepsilon''(\omega)
=1-\frac{\omega^{2}_\text{p}}{\Gamma^{2}_\text{e}+\omega^{2}}+
\text{i} \frac{ \Gamma_\text{e} \omega^{2}_\text{p}}{\omega (\Gamma^{2}_\text{e}+\omega^{2})},
\label{epsilon}
\end{split}
\end{equation}
where $\omega_\text{p}=\sqrt{e^{2}n_\text{e}/m_\text{e}\varepsilon_{0}}$ is the electric plasma frequency,
$\varepsilon_{0}$ is permittivity of free space,
$n_\text{e}$ is density of electrons,
$\text{e}$ is the electric charge,
$m_\text{e}$ is effective mass of electron and
$\varepsilon_{0}$ is the permittivity of free space.
In Eq.~(\ref{epsilon}$), \Gamma_\text{e}$ is electric damping constant and $\omega$ is the operating frequency.

To describe the magnetic response of metamaterials
we use the equivalent effective RLC circuits
(an electrical circuit consisting of
a resistor (R), an inductor (with inductance coefficient L), and a capacitor (C), connected in series or in parallel) \cite{penciu2010magnetic}.
Using this approach,
the magnetic permeability of metamaterials is in the form of Drude-Lorentz model \cite{penciu2010magnetic,cai2010optical}
\begin{equation}
\begin{split}
\frac{\mu(\omega)}{\mu_{0}}
&=\mu'(\omega)+\text{i} \mu''(\omega)\\
&=1-\frac{F \omega^{2} \omega^{2}_\text{p} (\omega^{2}-\omega^{2}_{0})}
{\Gamma^{2}_\text{m} \omega^{2}+(\omega^{2}-\omega^{2}_{0})^{2}}
+\text{i} \frac{ F \Gamma_\text{m} \omega^{3} \omega^{2}_\text{p}}{\Gamma^{2}_\text{m} \omega^{2}+(\omega^{2}-\omega^{2}_{0})^{2}}
\label{mu}
\end{split}
\end{equation}
with $\omega_{0}$ the magnetic resonance frequency,
$\Gamma_\text{m}$  magnetic damping constant,
$F$ magnetic oscillation strength
and $\mu_{0}$ the permeability of free space.

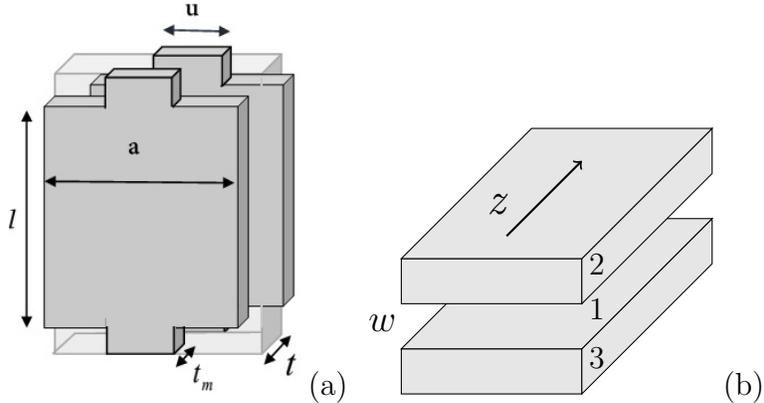
\begin{figure}
\centering
\includegraphics[width=.3\textwidth]{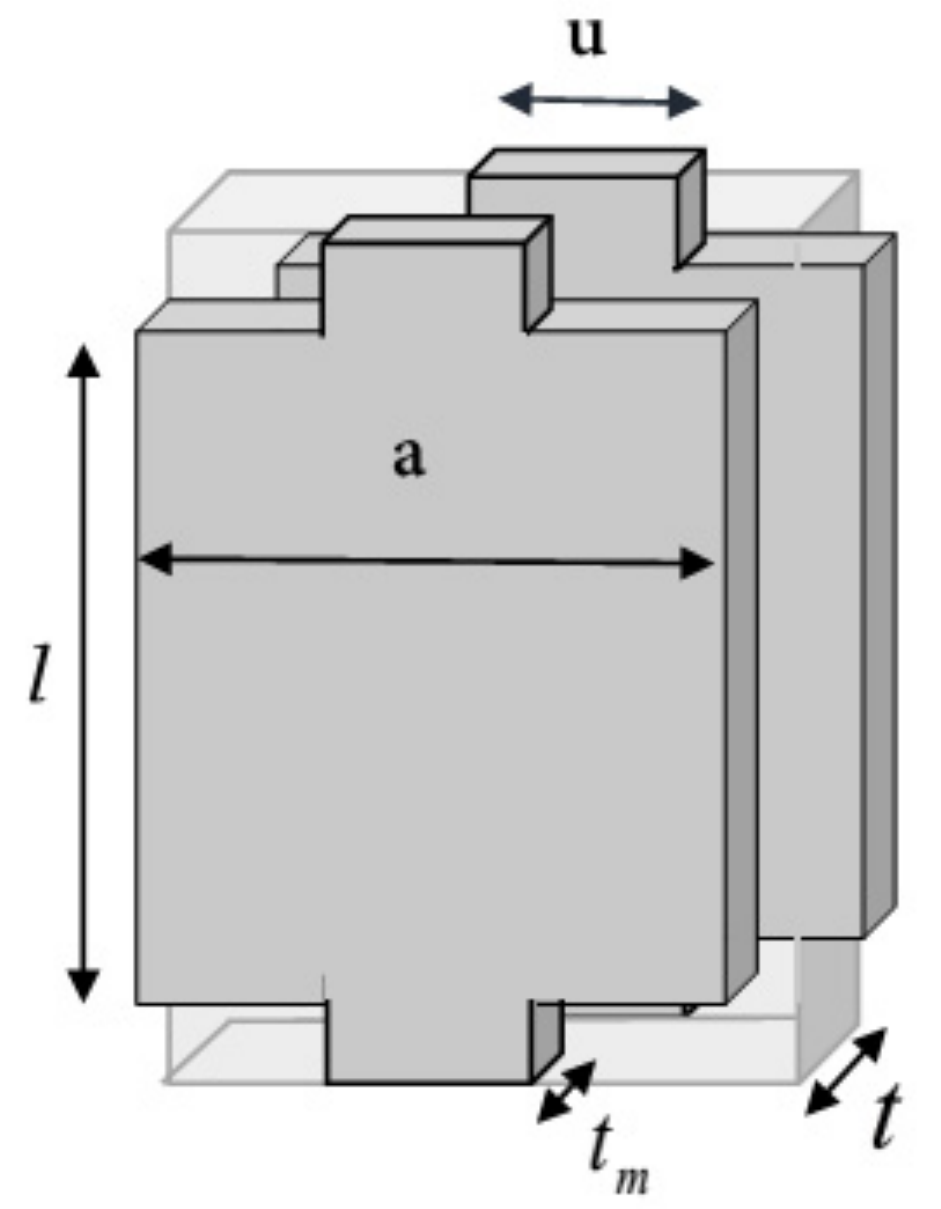}(a)
\begin{tikzpicture}

\begin{scope}[shift = {(0,1)}]
 
\pgfmathsetmacro{\cubex}{2.4}
\pgfmathsetmacro{\cubey}{.6}
\pgfmathsetmacro{\cubez}{4.5}

\node [below] at (19.35,.5) {\large $w$};

\draw[black,fill=gray!20] (22,-.1,0) -- ++(-\cubex,0,0) -- ++(0,-\cubey,0) -- ++(\cubex,0,0) -- cycle;
\draw[black,fill=gray!20] (22,-.1,0) -- ++(0,0,-\cubez) -- ++(0,-\cubey,0) -- ++(0,0,\cubez) -- cycle;
\draw[black,fill=gray!20] (22,-.1,0) -- ++(-\cubex,0,0) -- ++(0,0,-\cubez) -- ++(\cubex,0,0) -- cycle;

\draw[black,fill=gray!20] (22,1.1,0) -- ++(-\cubex,0,0) -- ++(0,-\cubey,0) -- ++(\cubex,0,0) -- cycle;
\draw[black,fill=gray!20] (22,1.1,0) -- ++(0,0,-\cubez) -- ++(0,-\cubey,0) -- ++(0,0,\cubez) -- cycle;
\draw[black,fill=gray!20] (22,1.1,0) -- ++(-\cubex,0,0) -- ++(0,0,-\cubez) -- ++(\cubex,0,0) -- cycle;

    \node [below] at (22.2,1.3) { $2$};
   \node [below] at (22.2,.71) { $1$};
   \node [below] at (22.2,.06) { $3$};

 \draw [->,thick] (21,1.4) -- (22,2.4);
 \node [below] at (20.9,2.1) {\Large $\bold{\it{z}}$};
   
\end{scope}
   \end{tikzpicture} (b)

\caption{(a) The fishnet unit-cell structure with $a$ the slabs width, $u$ the neck width, $l$ the
length of slabs and $t$ the separation between two slabs \cite{penciu2010magnetic} and (b) Slab metamaterial-dielectric waveguide with dielectric core (region 1) and metamaterial cladding (regions 2 and 3). Here $w$ is the waveguide width and $z$ is the propagation direction.}
\label{fishnetslab}
\end{figure}

In Eq.~(\ref{mu}),
$\Gamma_\text{m}$, $\omega_{0}$ and $F$
are functions of metamaterial structural parameters,
\begin{equation}
F=F'\frac{L}{L+L_{\rm e}}, \quad
F'=\frac{l t u}{V_{\rm uc}}
\label{F}
\end{equation}
is the ratio of the inter-pair volume to the unit-cell volume of metamaterial. In Eq.~(\ref{F}), $L$ is inductance, $L_{\text{e}}$ is kinetic inductance (both depend on metamaterial unit-cell parameters), $l$, $t$  and $u$ are the parameters of metamaterial unit-cell and $V_{\rm uc}$ is volume of the unit-cell (as shown in Fig.~\ref{fishnetslab}(a))~\cite{penciu2010magnetic}. Equation (\ref{F}) shows the dependence of $F$ on metamaterial unit-cell parameters. 

Magnetic resonance frequency and magnetic damping of the fishnet metamaterials
\begin{equation}
\omega_{0}=\frac{1}{\sqrt{{(L+L_{\rm e}})C}},\quad
\Gamma_{\text{m}}=\frac{R}{L+L_{\rm e}},
\label{omega0}
\end{equation}
depends on metamaterial unit-cell parameters (as in Fig.~\ref{fishnetslab}(a)).
Here, $C$ is the capacitance of the circuit and $R$ is the resistance of the system~\cite{penciu2010magnetic}.

The refractive index of metamaterials
\begin{equation}
\begin{split}
&n=n'+\text{i} n''
=\sqrt{\frac{\varepsilon \mu}{\varepsilon_{0} \mu_{0}}} \to\\
&n'=\frac{\sqrt {2}}{2 \sqrt {\varepsilon_{0} \mu_{0}}}\left(\sqrt {\sqrt {(\varepsilon' \mu'-\varepsilon'' \mu'')^{2}+(\varepsilon' \mu''+\varepsilon'' \mu')^{2}}+(\varepsilon' \mu'-\varepsilon'' \mu'')} \right),\\
&n''=\frac{\sqrt {2 \operatorname{sgn(\varepsilon' \mu''+\varepsilon'' \mu')}}}{2 \sqrt {\varepsilon_{0} \mu_{0}}}\left(\sqrt {\sqrt {(\varepsilon' \mu'-\varepsilon'' \mu'')^{2}+(\varepsilon' \mu''+\varepsilon'' \mu')^{2}}-(\varepsilon' \mu'-\varepsilon'' \mu'')} \right),
\label{n}
\end{split}
\end{equation}
consists of real and imaginary parts of both permittivity and permeability terms
, which depend on the metamaterial structural parameters.
Equations (\ref{epsilon})-(\ref{n})
express the effects of metamaterial structural parameters
on the EM susceptibilities of metamaterials.

We consider the structural parameters to vary in the acceptable ranges presented in literature~\cite{penciu2010magnetic, kamli2008coherent}. 
The magnetic oscillation strength
varies in the range $0.1-1$~\cite{penciu2010magnetic},
magnetic damping in the range of $10^{-5}\Gamma_\text{e} < \Gamma_\text{m} < \Gamma_\text{e}$~\cite{kamli2008coherent}
and magnetic resonance frequency in the range of $0.1 \omega_\text{e}-0.9 \omega_\text{e}$.
Therefore the effective parameter space (containing acceptable variation range of $F$, $\omega_{0}$ and $\Gamma_\text{m}$) is a three-dimensional real space.
From this three-dimensional parameter space,
we map to a two-dimensional complex space (for specific parameter choice).
By choosing examples in the considered parameter space,
we then characterize the effects of these parameters on
the EM characteristics of metamaterial waveguides. 

We also investigate robustness analysis of the systems affected by Gaussian errors on metamaterial structural parameters. 
A robustness analysis will allow us to determine the stability of the systems when errors in the structural parameters are present. Such errors can occur as a natural by-product of the metamaterial fabrication process or due to temperature fluctuations, for example.
 
\subsection{Slab Waveguide}

In this subsection
we present the characteristics of slab waveguides
and review the modal behavior of slab metamaterial-dielectric waveguides.
Recent studies on modes' behavior EM waveguides are based on
waveguides with different geometries and different materials \cite{iizuka2002elements}
and the applications of metamaterials in waveguides have attracted interest \cite{sang2015electromagnetic,lavoie2012low,ruppin2001surface}.
In this paper,
we study slab metamaterial-dielectric waveguide,
with dielectric core and metamaterial cladding
where the core with the width of $w=4\pi c/\omega_\text{e}$
is sandwiched by two metamaterial layers (Fig.~\ref{fishnetslab}(b)).

Slab metamaterial-dielectric waveguides
can support transverse magnetic (TM)
and transverse electric (TE) modes \cite{sang2015electromagnetic,tong2014advanced}
with three mode behaviors
namely ordinary,
surface-plasmon polariton (SPP) \cite{ruppin2001surface}
and hybrid ordinary-SPP modes, which we simply call hybrid modes~\cite{lavoie2012low}.
More specifically,
hybrid modes are the product of two features in the transverse direction:
evanescent (SPP modes) and oscillatory (ordinary modes) features \cite{tong2014advanced,lavoie2012low}.
We determine the modal behavior of the waveguide by
comparing the real and imaginary parts of the wave-number
perpendicular to the waveguide propagation direction ($\rm Re(\gamma_{1})/\rm Im(\gamma_{1})$) \cite{pendry2006controlling}.
The wave-number is obtained by numerically
solving the waveguide modal dispersion equation.
We focus on hybrid modes region,
as this mode is one of the significant properties of
metamaterial-dielectric waveguides.
The condition for hybrid mode existence is $10^{-2}\leq \rm Re(\gamma_{1})/\rm Im(\gamma_{1})\leq 50$~\cite{lavoie2012low}.

The dispersion equation for TM modes of a slab guide is \cite{lavoie2012low} 
\begin{equation}
\frac{\varepsilon_{1}}{\gamma_{1}} \left(\frac{\varepsilon_{2}}{\gamma_{2}}+\frac{\varepsilon_{3}}{\gamma_{3}}\right)=-\left(\frac{\varepsilon^{2}_{1}}{\gamma^{2}_{1}}+\frac{\varepsilon_{2}\varepsilon_{3}}{\gamma_{2}\gamma_{3}}\right) \tanh(\gamma_{1}w)
\label{6}
\end{equation}
with $\gamma_{j}=\sqrt{k^{2}_{z}-\omega^{2}\varepsilon_{j}\mu_{j}}$ the complex wave number for the transverse component of the field,
$j=1,2,3$ refer to the core and two cladding layers, respectively \cite{lavoie2012low},
$k_{z}$ the propagation constant and
$w$ the core width.
To limit the number of metamaterial parameters we consider the symmetric waveguide, in which regions 2 and 3 are the same
($\varepsilon_{2}=\varepsilon_{3}$  and $\gamma_{2}=\gamma_{3}$). 

We investigate the effects of metamaterial tailoring
on hybrid mode behavior in metamaterial-dielectric waveguides.
To study the modes' behavior,
we compare the relative size of real and imaginary parts of $\gamma_{1}$. 

\section{Results and discussion}

In this section, we investigate the effect of metamaterial structural parameters
on the behavior of refractive index in metamaterials.
Furthermore,
we analyze the robustness
of metamaterials
under fluctuations in structural parameters.
We then analyze the behavior of hybrid modes in slab metamaterial-dielectric waveguides
by varying metamaterial structural parameters.

\subsection{Metamaterials}

\begin{figure}[h]
\centering
\includegraphics[width=0.34\textwidth]{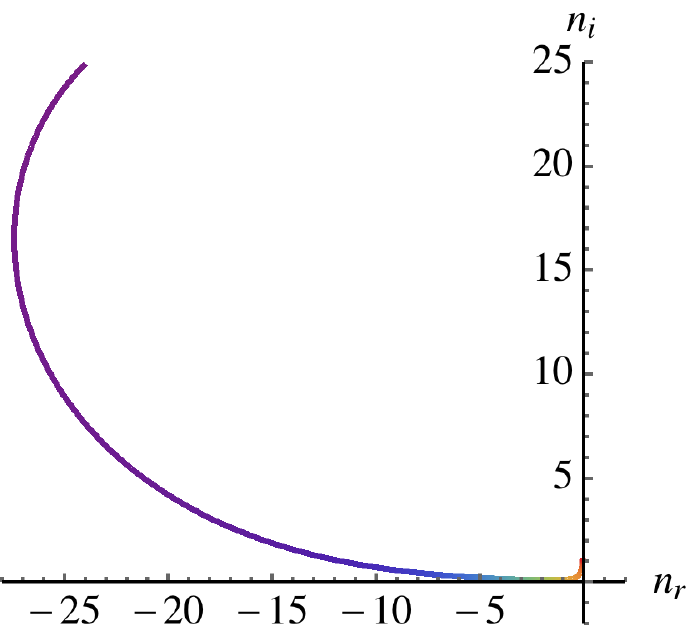}
\includegraphics[width=0.058\textwidth]{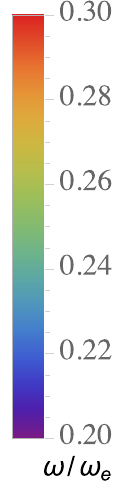}(a)
\includegraphics[width=0.38\textwidth]{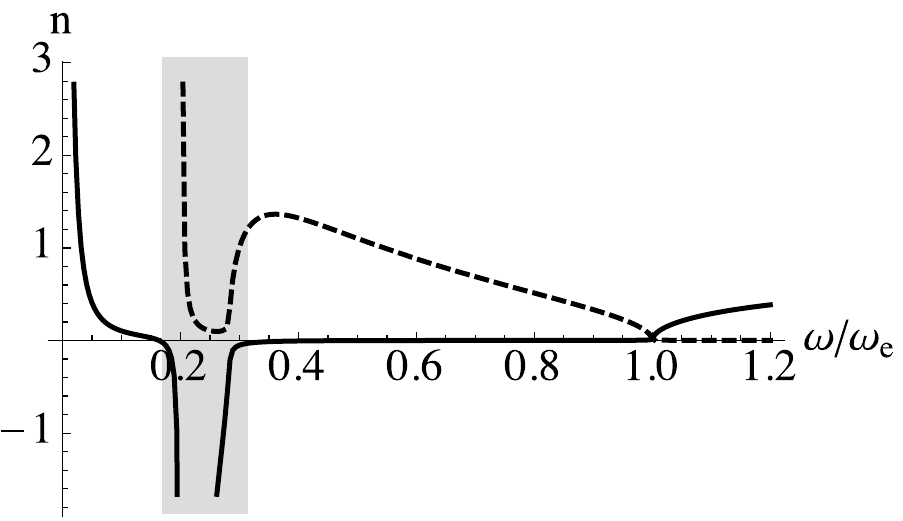}(b)
\caption{(a) The complex graph for real and imaginary parts of refractive index
by varying the operating frequency,
the legend shows the variation of operating frequency at frequencies around negative refractive index frequency region
starting from $\omega\approx0.2 \omega_{\rm e}$(violet) to $w\approx0.3 \omega_{\rm e}$(red),
(b) The plot of real part (solid lines) and imaginary part (dashed lines) of refractive index
of the metamaterial as a function of frequency,
the shaded region represents negative refractive index region.
The metamaterial parameter values used are:
$\omega_\text{e}=1.37\times 10^{16} \text{s}^{-1}$,
$\omega_{0}=0.2\omega_\text{e}$,
$F=0.5$ and
$\Gamma_\text{m}=\Gamma_\text{e}=2.73\times 10^{13} \text{s}^{-1}$~\cite{lavoie2012low}.}
\label{nrniw}
\end{figure}

From the three-dimensional parameter space ($F$, $\omega_{0}$ and $\Gamma_{\text{m}}$) of metamaterial structural parameters,
we map the refractive index to a two-dimensional complex plane (Fig.~\ref{nrniw}(a))
for a set of parameters presented in the caption of Fig.~\ref{nrniw}(a).
From this complex plane,
we choose a point, for instance, 
to investigate the behavior of real and imaginary parts of refractive index
as a function of the operating frequency (Fig.~\ref{nrniw}(b)).

Figure \ref{nrniw}(b) presents the variation of real and imaginary parts of refractive index
as a function of operation frequency.
From Fig.~\ref{nrniw}(b) we see that
the real part of refractive index becomes negative
in the frequency region $0.2\omega_\text{e}\leq \omega \leq 0.3\omega_\text{e}$ (shaded region).

\begin{figure}
\centering
\includegraphics[width=0.33\textwidth]{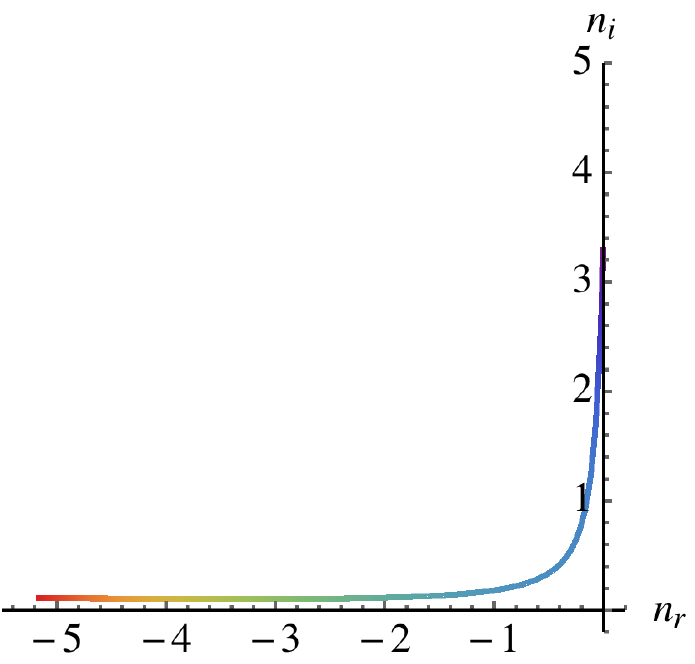}
\includegraphics[width=0.08\textwidth]{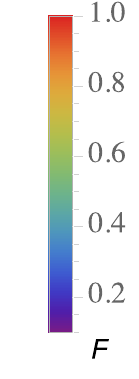}(a)
\includegraphics[width=0.35\textwidth]{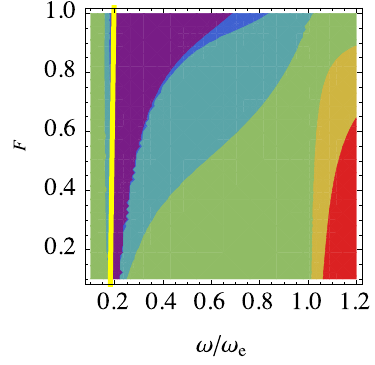}
\includegraphics[width=0.065\textwidth]{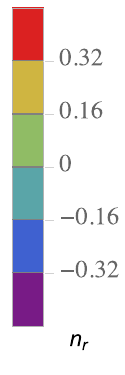}(b)
\includegraphics[width=0.42\textwidth]{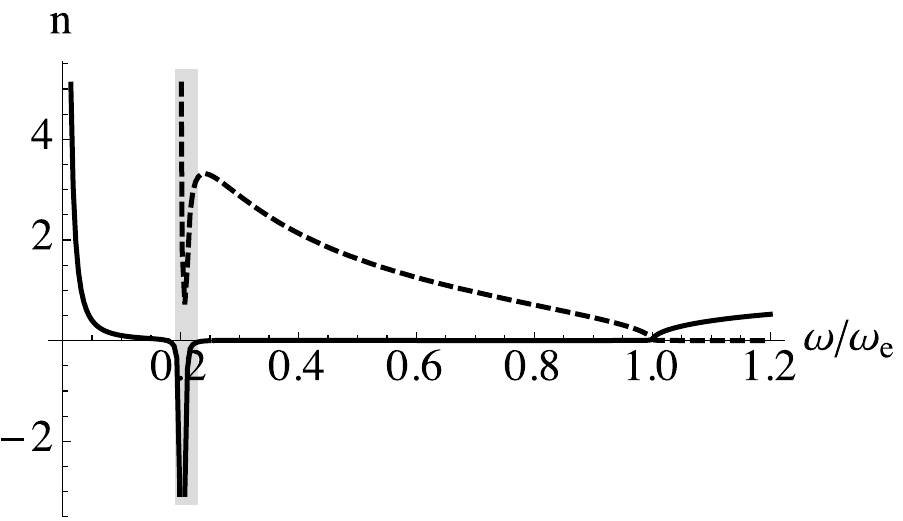}(c)
\includegraphics[width=0.42\textwidth]{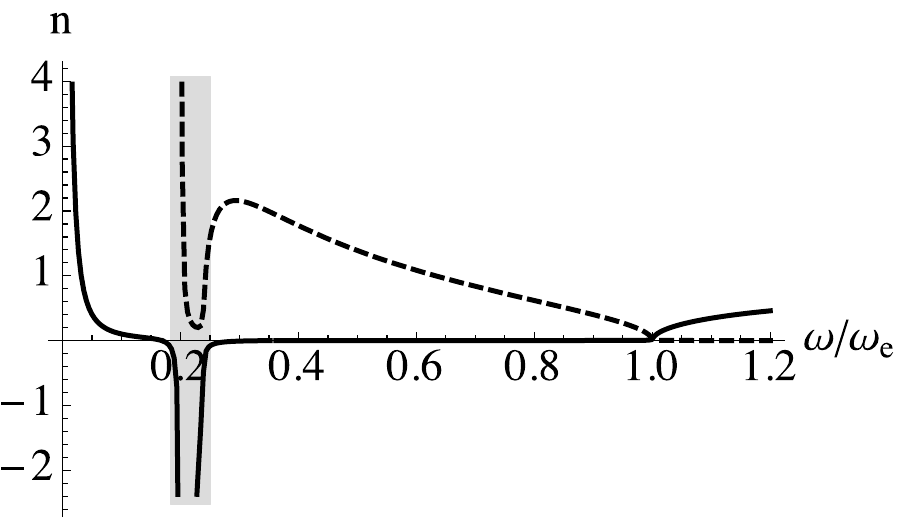}(d)
\caption{(a) The plot of real and imaginary parts of refractive index by changing $F$
using the set of metamaterial parameters as Fig.~\ref{nrniw} at the frequency $\omega=0.25 \omega_\text{e}$ (at negative refractive index frequency region). The Legend shows the variation of $F$ at the range of $0.1-1$.
(b) The contour plot for real part of refractive refractive index of the metamaterial
as a function of frequency and magnetic oscillation strength
using the set of metamaterial parameters as Fig.~\ref{nrniw}. The legend presents the variation of the real part of refractive index and the yellow line shows the minimum of refractive index.
Plots (c) and (d) represent the behavior of real and imaginary parts of refractive index
for the same set of parameters
with F=0.1 and F=0.3, respectively.}
\label{3DF}
\end{figure}

\begin{figure}[h]
\centering
\includegraphics[width=0.34\textwidth]{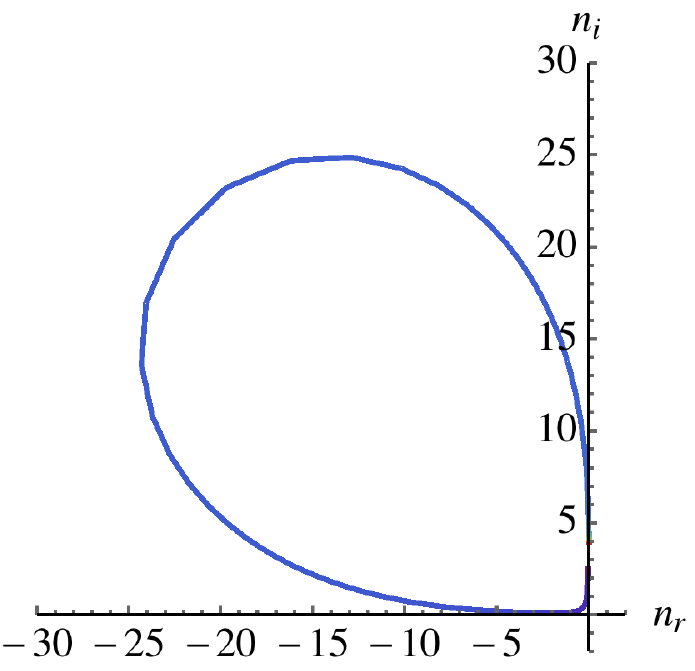}
\includegraphics[width=0.08\textwidth]{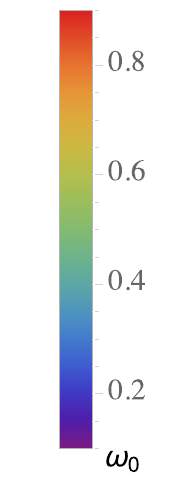}(a)
\includegraphics[width=0.36\textwidth]{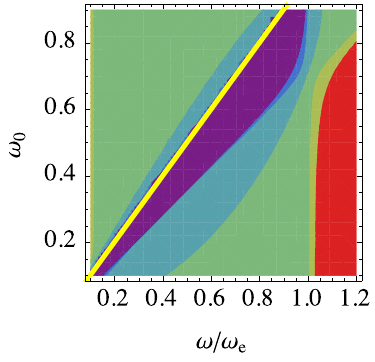}
\includegraphics[width=0.076\textwidth]{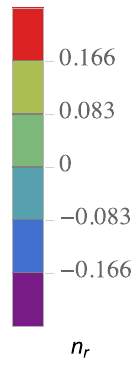}(b)
\includegraphics[width=0.42\textwidth]{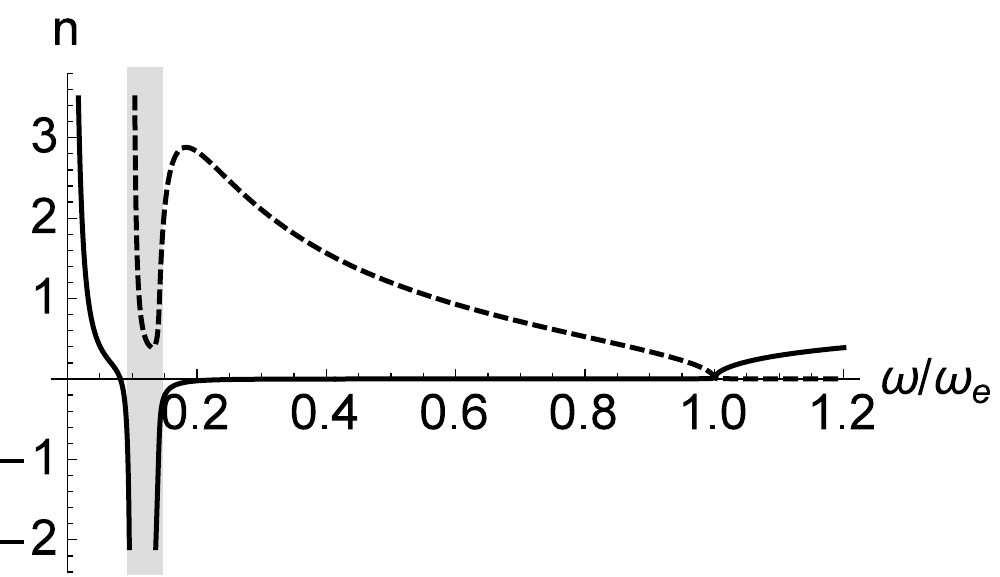}(c)
\includegraphics[width=0.42\textwidth]{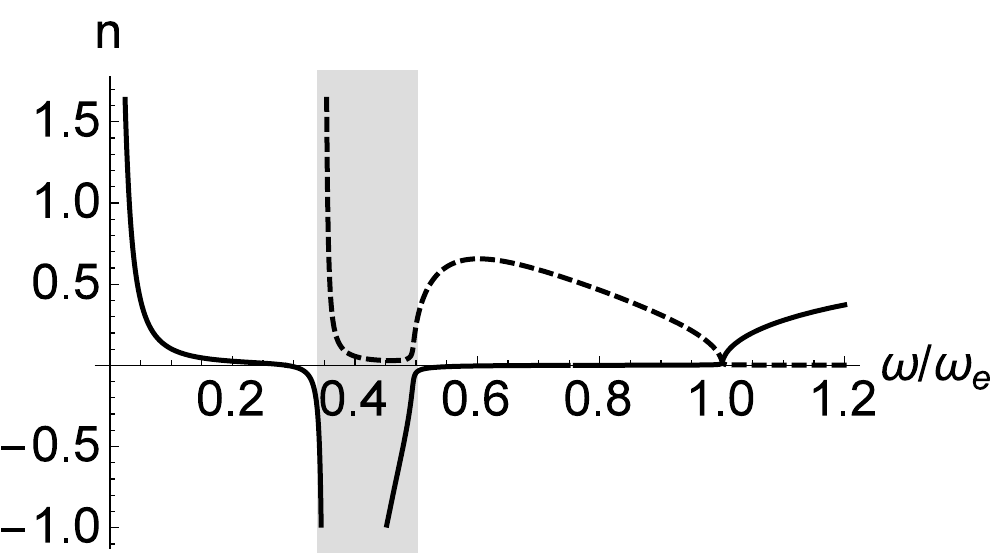}(d)
\caption{The plot of
(a) real and imaginary part of refractive index by varying $\omega_{0}$
using the set of metamaterial parameters as Fig.~\ref{nrniw} at $\omega=0.25\omega_{\rm e}$ (at negative refractive index frequency region), the legend shows the variation of $\omega_{0}$ at the range of $0.1 \omega_{\rm e}-1 \omega_{\rm e}$ and
(b) real part of refractive index
of the metamaterial as a function of frequency and magnetic resonance frequency
using the set of metamaterial parameters as Fig.~\ref{nrniw}. The legend presents the variation of the real part of refractive index and the yellow line shows the minimum of refractive index.
Plots (c) and (d) represent the behavior of real and imaginary parts of refractive index
for the same set of parameters
with $\omega_{0}=0.1\omega_\text{e}$ and
$\omega_{0}=0.35\omega_\text{e}$, respectively.}
\label{3Dw0}
\end{figure}

\begin{figure}
\centering
\includegraphics[width=0.29\textwidth]{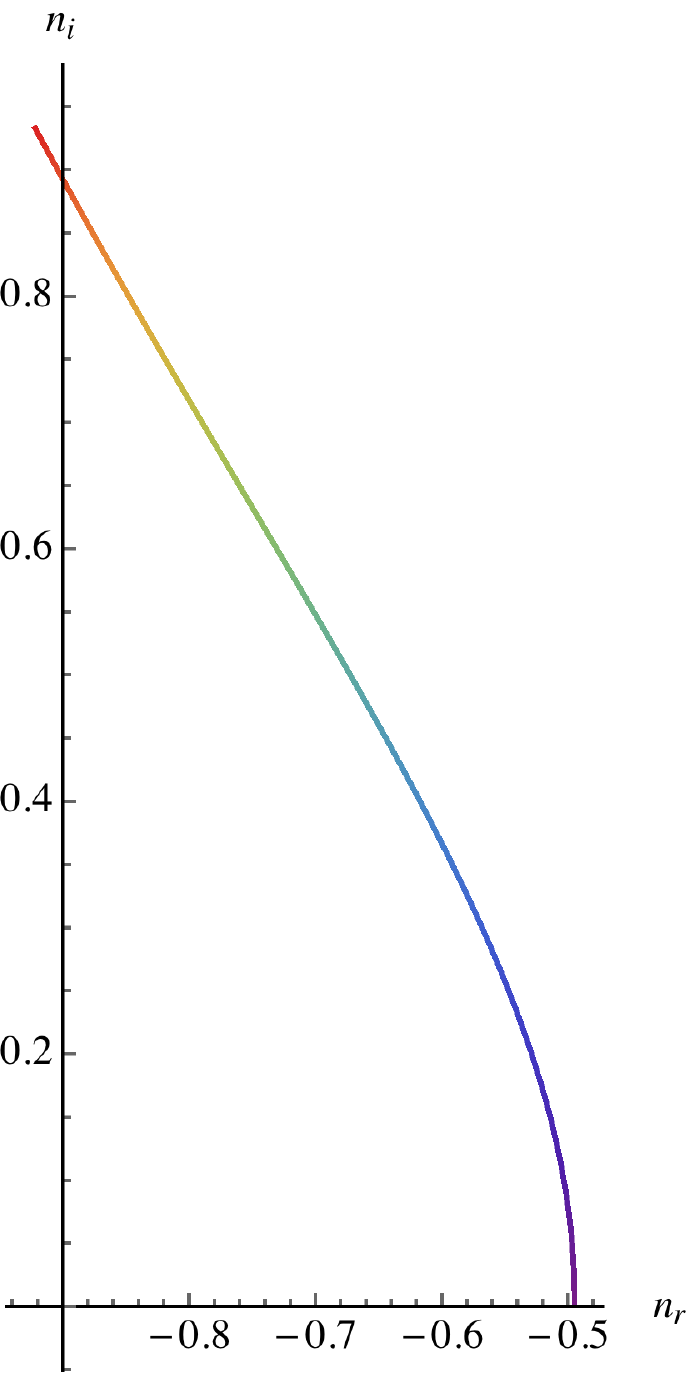}
\includegraphics[width=0.08\textwidth]{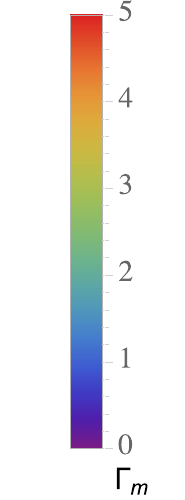}(a)
\includegraphics[width=0.36\textwidth]{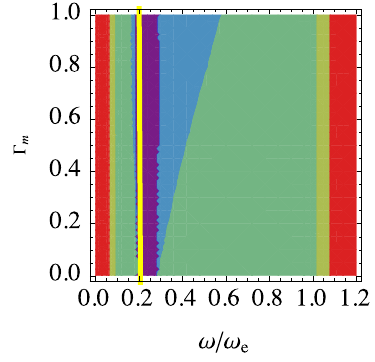}
\includegraphics[width=0.072\textwidth]{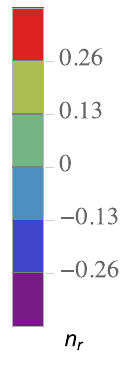}(b)
\includegraphics[width=0.42\textwidth]{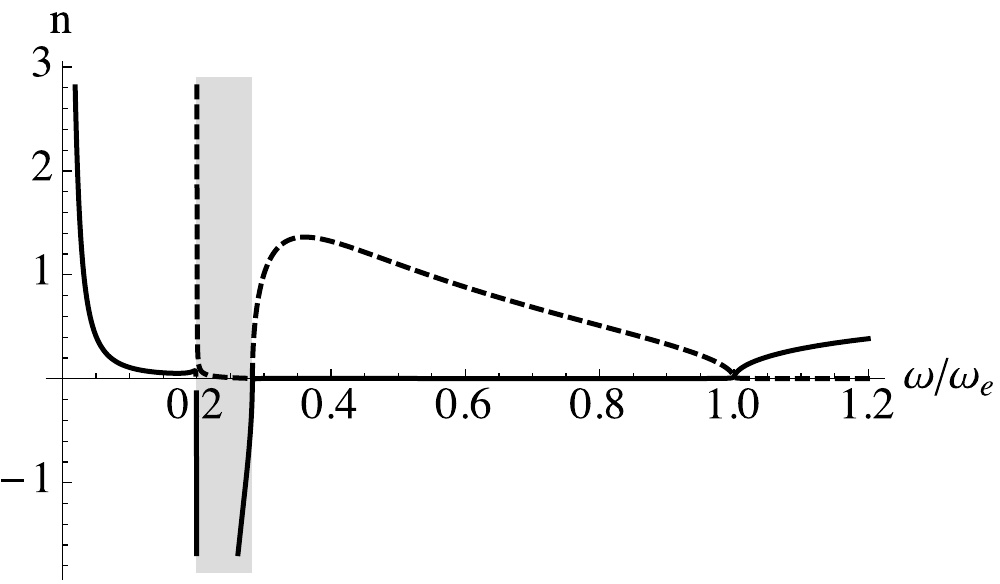}(c)  
\includegraphics[width=0.42\textwidth]{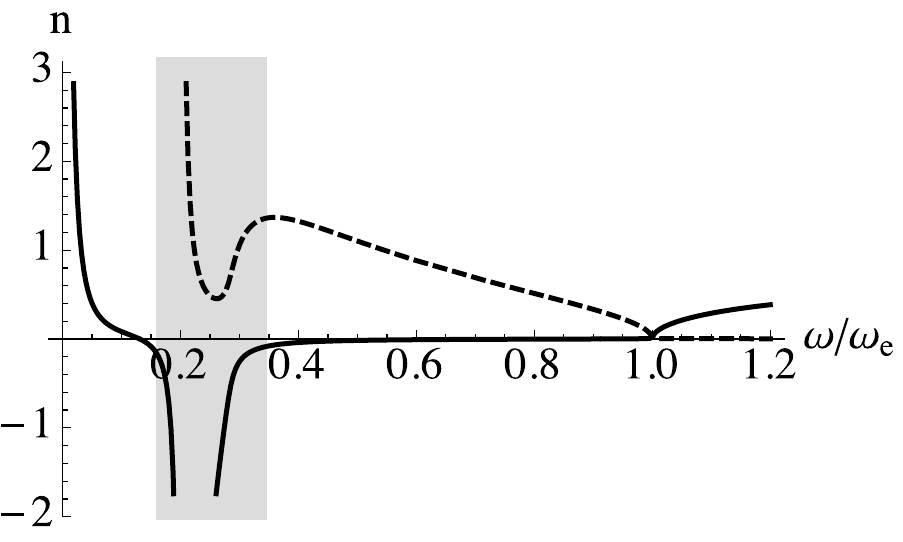}(d)
\caption{The plot of
(a) real and imaginary part of refractive index by varying $\Gamma_\text{m}$ using the set of metamaterial parameters as Fig.~\ref{nrniw} at $\omega=0.28 \omega_\text{e}$, the legend shows the variation of $\Gamma_\text{m}$ at the range of $0.001-1\Gamma_\text{e}$ and
(b) real part and of refractive index of the metamaterial
as a function of frequency and magnetic damping
using the set of metamaterial parameters as Fig.~\ref{nrniw}. The legend presents the variation of the real part of refractive index and the yellow line shows the minimum of refractive index.
Plots (c) and (d) represents the behavior of real and imaginary parts of refractive index
for $\Gamma_\text{m}=0.01\Gamma_\text{e}$ and $\Gamma_\text{m}=5\Gamma_\text{e}$, respectively.}
\label{3Dgm}
\end{figure}

By considering different values for structural parameters in the three-dimensional parameter space
and mapping to the two-dimensional complex plane,
we investigate the behavior of refractive index
as illustrated in Figs.~\ref{3DF}-\ref{3Dgm}.
We choose the parameter values in the permitted ranges
to investigate the variations in EM responses of metamaterials and modes in the waveguides
by varying metamaterial structural parameters.
In Fig.~\ref{3DF}(a) the variation of real and imaginary parts of refractive index is presented as a function of $F$ at the frequency $\omega=0.25 \omega_\text{e}$.
By increasing $F$ from $0.1$ to $0.6$, the real part of refractive index becomes more negative values and the imaginary part increases.
From this two-dimensional complex plane presented in Fig.~\ref{3DF}(a),
we investigate the variation of real
part of refractive index by changing $F$ and $\omega$ (Fig.~\ref{3DF}).
For a more clear understanding of the behavior,
Figs.~\ref{3DF}(c) and (d) present the real and imaginary parts of the refractive index
for specific values of $F=0.1$ and $F=0.3$, as examples,
by changing the operating frequency. 

In Figs.~\ref{3DF}, the larger negative values of refractive index are illustrated as dark violet regions.
By increasing the magnetic oscillation strength, as presented in Figs.~\ref{3DF},
the negative refractive index region increases.
This change of $n<0$ frequency region manifests the sensitiveness of metamaterial refractive index on the magnetic oscillation strength.  
Therefore, to achieve wider negative index regions,
larger values of $F$ are required.
Conversely, if more sharp frequency region is required for negative refractive index,
smaller $F$ is needed.

We investigate the effect of magnetic resonance frequency
on the refractive index of metamaterials by considering
different values of $\omega_{0}$ and fixed $F$ and $\Gamma_\text{m}$ in three-dimensional parameter space.
Figure \ref{3Dw0}(a) presents the complex plane for
variation of real and imaginary parts of the refractive index
by changing $\omega_{0}$.
By varying $\omega_{0}$ from $0.1 \omega_\text{e}$ to $0.5 \omega_\text{e}$,
the real and imaginary parts of refractive index vary as Fig.~\ref{3Dw0}(a).
The contour plots for the behavior of refractive index by varying $\omega_{0}$ and $\omega$
is shown in Fig.~\ref{3Dw0}(b).
Figures ~\ref{3Dw0}(c) and (d) show the change in refractive index
for $\omega_{0}=0.1 \omega_\text{e}$ and $\omega_{0}=0.35 \omega_\text{e}$, as examples,
at different operating frequencies.
By increasing $\omega_{0}$,
the real and imaginary parts of refractive index
shift to higher frequencies, as illustrated in Fig.~\ref{3Dw0}.
For $\omega_{0}=0.1\omega_\text{e}$,
the negative refractive index region is $0.1\omega_\text{e}\leq\omega\leq0.15\omega_\text{e}$
and increasing $\omega_{0}$ to $0.35\omega_\text{e}$ shifts
this negative refractive index region to $0.35\omega_\text{e}\leq\omega\leq0.5\omega_\text{e}$.
Therefore, increasing $\omega_{0}$
shifts the negative refractive index region
to higher frequencies. 

The next parameter is magnetic damping constant
whose effect on the negative-index region is illustrated in Figs.~\ref{3Dgm}.
Figure \ref{3Dgm}(a) represents the variation of real and imaginary parts of refractive index
by changing magnetic damping at $\omega=0.28 \omega_\text{e}$.
We consider a different frequency, than the previous parameters, for investigating the effect of $\Gamma_\text{m}$ because the effects of changing $\Gamma_\text{m}$ is more obvious at this frequency. 
By decreasing $\Gamma_\text{m}$
the real part of refractive index gets less negative values
where the imaginary part becomes smaller (reaches to zero at very small values of $\Gamma_\text{m}$).
Figure \ref{3Dgm}(b) verifies that
by increasing $\Gamma_\text{m}$,
the real part of refractive index gets more negative values.
To see the effect of $\Gamma_\text{m}$
on the imaginary part of refractive index
we refer to Fig \ref{3Dgm}(c),
which is the variation of real and imaginary parts of refractive index
by varying frequency for $\Gamma_\text{m}=0.01\Gamma_\text{e}$.
Figure \ref{3Dgm}(c) shows that
by decreasing magnetic damping constant,
the imaginary part of refractive index reduces
where, absorption is very small for
$0.2\omega_\text{e}\leq\omega\leq0.3\omega_\text{e}$
frequency region by decreasing $\Gamma_\text{m}$ to $\Gamma_\text{m}=0.01\Gamma_\text{e}$.

In this subsection, we presented a brief overview
for the effects of metamaterial structural parameters on
the EM response of these materials.
Our results demonstrate the fact that
three effective metamaterial structural parameters
generate a three-dimensional parameter space,
from which we map to two-dimensional complex plane for the refractive index of metamaterials.
As our results present, each point in the three-dimensional parameter space
corresponds to a specific EM response from metamaterials
that enables the user-intended applications of these structures.

The problem with each point in the three-dimensional parameter space is that
metamaterials construction doesn't allow the consideration of exact value for
each of three structural parameters of metamaterials.
Such a restriction requires considering a three-dimensional volume (with permitted fluctuations in each of the parameters) around each point in space, instead of considering an exact point,
to be more realistic.
Considering the parameters fluctuations
require more analysis to show the robustness of the metamaterial systems to
errors in their structures. 

\subsection{Robustness Analysis}

Robustness of a system is its ability
to tolerate perturbations of the input parameters~\cite{Robustbook}.
Examining the robustness of metamaterials under
fluctuations of their structural parameters
gives us a sense of the ability of metamaterials to resist fluctuations
in the metamaterial structural parameters.
We examine the robustness of the metamaterials
by applying Gaussian errors
on metamaterial structural parameters
in the three-dimensional parameter space,
consisting of $\omega_{0}$, $F$ and $\Gamma_\text{m}$.

\begin{figure}
\centering
\includegraphics[width=0.46\textwidth]{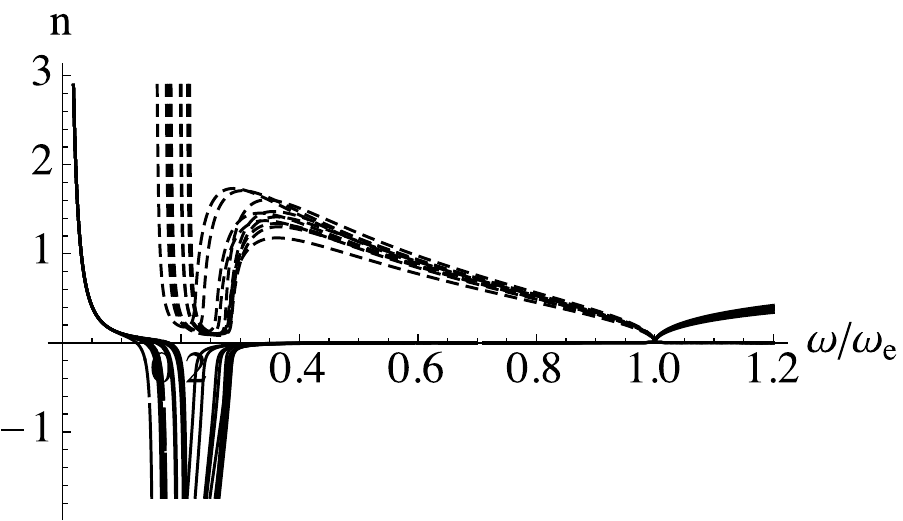}
\caption{Plot of real part (solid line) and imaginary part (dashed line) of refractive index
as a function of frequency
by applying 10\% errors with Gaussian distribution
on metamaterial structural parameters $\omega_{0}$, $F$ and $\Gamma_\text{m}$ (using parameter values as Fig.~\ref{nrniw}).}
\label{nrnirobust}
\end{figure}

\begin{figure}
\centering
\includegraphics[width=0.36\textwidth]{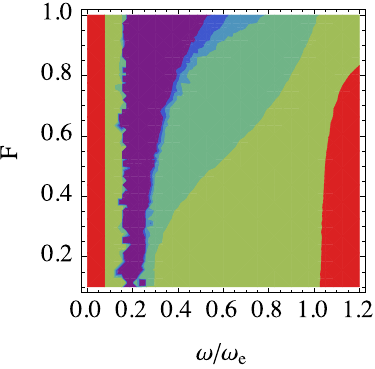}
\includegraphics[width=0.084\textwidth]{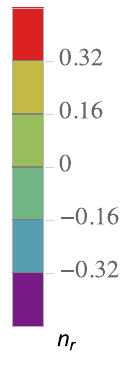}(a)
\includegraphics[width=0.36\textwidth]{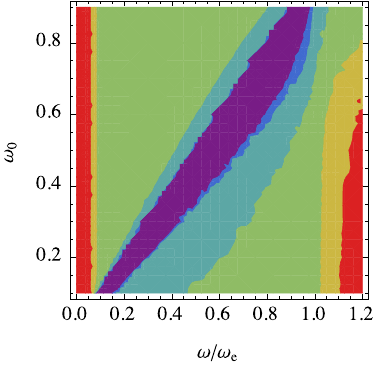}
\includegraphics[width=0.08\textwidth]{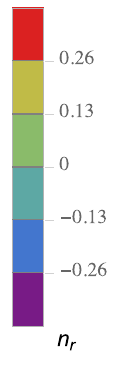}(b)
\includegraphics[width=0.36\textwidth]{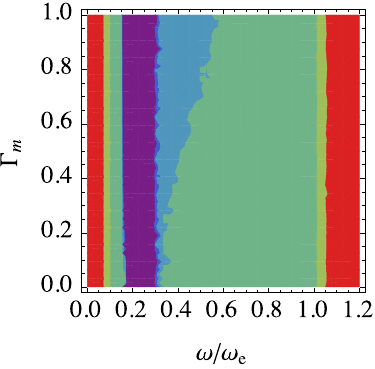}
\includegraphics[width=0.08\textwidth]{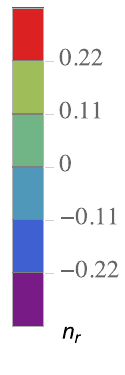}(c)
\caption{Plot of real part of refractive index by applying 10\% errors with Gaussian distribution on (a) $\omega_{0}$ and $\Gamma_\text{m}$, (b) on $F$ and $\Gamma_\text{m}$ and (c) on $\omega_{0}$ and $F$.}
\label{nrFw0gmrobust}
\end{figure}

Figure \ref{nrnirobust} shows the variation of
metamaterial refractive index as a function of frequency
by adding 10\% Gaussian fluctuations on the structural parameters.
By applying Gaussian errors to the input parameters,
the functional form of metamaterial refractive index doesn't change
(compare Figs.~\ref{nrniw} and \ref{nrnirobust}).
The robust response to the Gaussian error in the input parameters
demonstrates the ability of metamaterials to resist change without adapting the initial parameters.
This robustness is more evident in Figs.~\ref{nrFw0gmrobust}(a)-(c).
Figure \ref{nrFw0gmrobust}(a) shows the EM response of metamaterials
(real part of refractive index)
to the Gaussian fluctuations on $\omega_{0}$ and $\Gamma_\text{m}$.
By applying the Gaussian errors on $F$ and $\Gamma_\text{m}$, the variation of refractive index with respect to $\omega_{0}$ and $\omega$ is shown in \ref{nrFw0gmrobust}(b).
Figure \ref{nrFw0gmrobust}(c) shows the variation of refractive index
with respect to $\Gamma_\text{m}$ and $\omega$
by applying fluctuations on $F$ and $\omega_{0}$.
By comparing Figs.~\ref{nrFw0gmrobust}(a)-(c) with 
Figs.~\ref{3DF}, \ref{3Dw0} and \ref{3Dgm},
we see the robustness of metamaterials to the Gaussian errors
on the input structural parameters.

We examined the robustness of metamaterials
by considering Gaussian fluctuations in
metamaterial structural parameters.
Our results demonstrate the robustness of metamaterials to
the inaccuracy in structural parameters up to the level of almost 10\% fluctuations in each parameter in the three-dimensional parameter space,
which are the facts of realistic metamaterials~\cite{Khurgin2015}.

\subsection{Metamaterial Waveguides}

By considering the effect of metamaterials structural parameters
on the EM properties of metamaterials,
we now investigate the influence of metamaterial tailoring
on the modes' behavior in metamaterial-dielectric waveguides.
The supported modes of a waveguide depend on a number of parameters
including waveguides geometry, material, and operating frequency~\cite{wang2007nanoscale}.
Previous work shows that
hybrid modes exist in metamaterial-dielectric waveguides
in the region that $10^{-2}\leq \rm Re(\gamma_{1})/\rm Im(\gamma_{1})\leq 50$~\cite{lavoie2012low}.
We use this property of hybrid modes to investigate their behavior
under tailoring metamaterials EM responses. 

\begin{figure}
\centering
\includegraphics[width=0.48\textwidth]{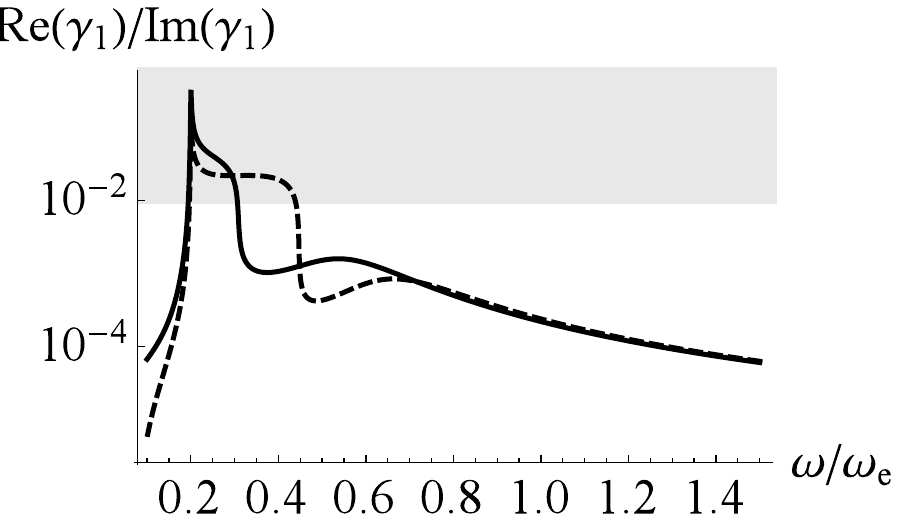}
\caption{Plot of the $\rm Re(\gamma_{1})/\rm Im(\gamma_{1})$
for $\rm TM_{0}$ (solid line) and $\rm TM_{1}$ (dashed line) modes
using the set of parameters presented in Fig.~\ref{nrniw}.
The shaded region corresponds to hybrid-mode region.}
\label{gmguide}
\end{figure}

We consider $\rm TM_{0}$ and $\rm TM_{1}$ modes
as examples to study the behavior of modes in the waveguide
by tailoring metamaterial structure.
The behavior of real to imaginary parts of $\gamma_{1}$ for metamaterial-dielectric slab waveguide for the set of parameters as Fig.~\ref{nrniw} is presented in Fig.~\ref{gmguide}.
Figure \ref{gmguide} shows that the hybrid modes exist in the
frequency region
$0.2\omega_\text{e}\leq\omega\leq0.3\omega_\text{e}$ for $\rm TM_{0}$
and $0.2\omega_\text{e}\leq\omega\leq0.45\omega_\text{e}$ for $\rm TM_{1}$ modes.  

\begin{figure}
\centering
\includegraphics[width=0.42\textwidth]{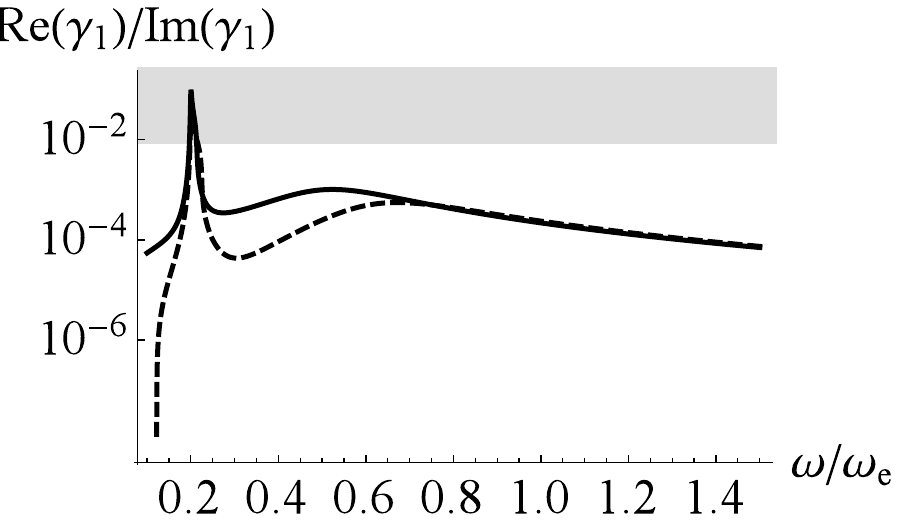}(a)
\includegraphics[width=0.42\textwidth]{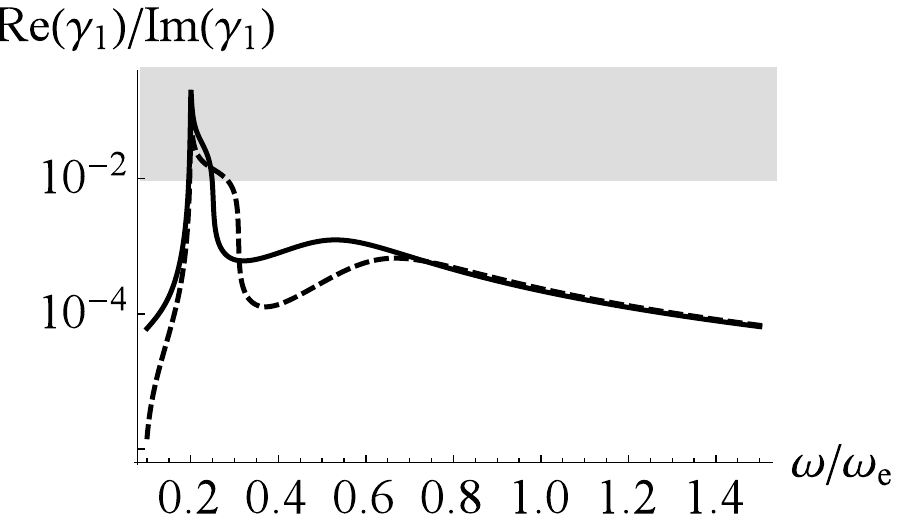}(b)
\includegraphics[width=0.42\textwidth]{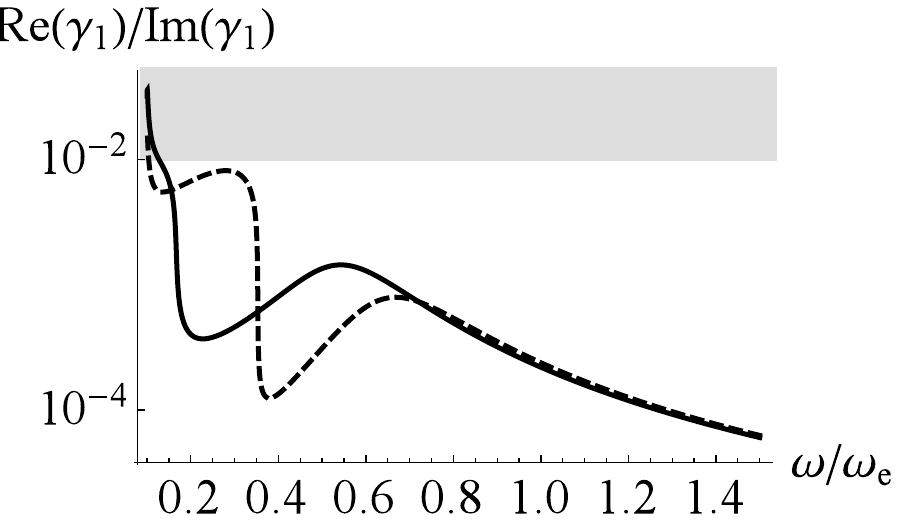}(c)
\includegraphics[width=0.42\textwidth]{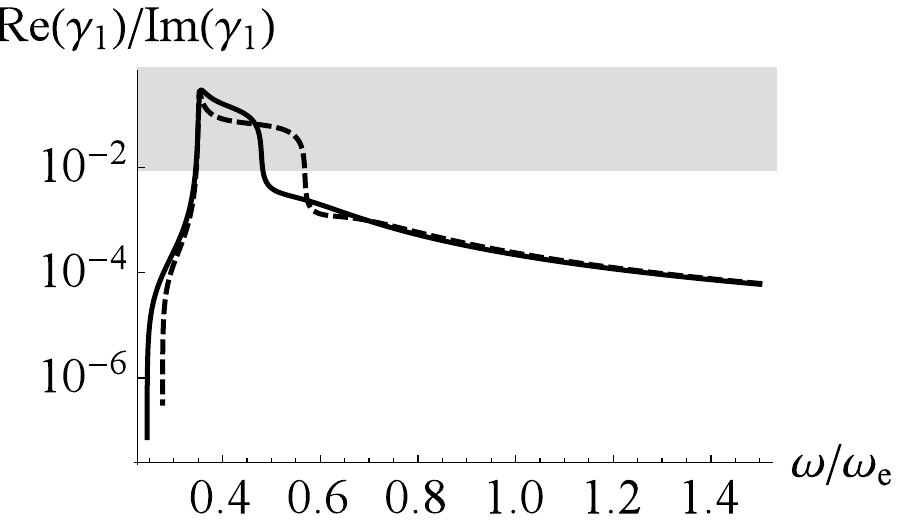}(d)
\includegraphics[width=0.42\textwidth]{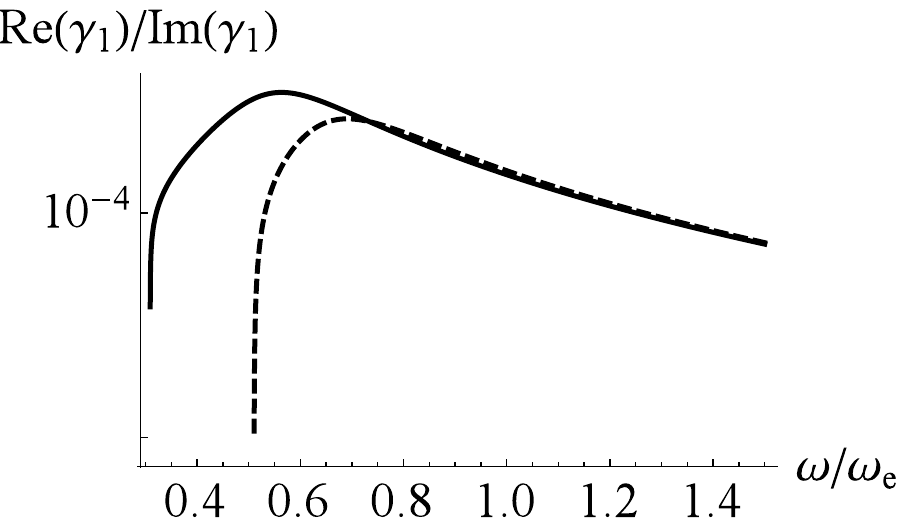}(e)
\caption{Plot of the $\rm Re(\gamma_{1})/\rm Im(\gamma_{1})$ for
$\rm TM_{0}$ (solid line) and $\rm TM_{1}$ (dashed line) modes
for (a) $F=0.1$,
(b) $F=0.3$,
(c) $\omega_{0}=0.1\omega_\text{e}$,
(d) $\omega_{0}=0.35\omega_\text{e}$,
and (e) $\Gamma_\text{m}=0.1\Gamma_\text{e}$.
The shaded region corresponds to hybrid-mode region.}
\label{hybridmodechange}
\end{figure}

We now investigate the effect of variations in metamaterials structural parameters
on the hybrid modes' behavior
in the metamaterial-dielectric waveguide.
Figures \ref{hybridmodechange}(a)-(e) illustrate the behavior of $\rm TM_{0}$ and $\rm TM_{1}$ modes
in metamaterial-dielectric slab waveguide by considering different values for $F$, $\omega_{0}$ and $\Gamma_\text{m}$.
The hybrid mode region
for $\rm TM_{0}$ and $\rm TM_{1}$ modes
increases by increasing $F$, as illustrated in Figs.~\ref{hybridmodechange}(a) and (b).

The effect of varying $\omega_{0}$ on the behavior of hybrid modes is shown in Figs.~\ref{hybridmodechange}(c) and (d),
that show
the shift of hybrid modes frequency region,
for $\rm TM_{0}$  and $\rm TM_{1}$  modes,
to higher frequencies with increasing magnetic resonance frequency. 
By decreasing $\gamma_\text{m}$, the behavior of both $\rm TM_{0}$ and $\rm TM_{1}$ modes changes
such that no hybrid modes exist for both $\rm TM_{0}$ and $\rm TM_{1}$ mode (Fig.~\ref{hybridmodechange}(e)).
Therefore, the existence of damping for metamaterials
in the waveguides cladding
is required for having hybrid modes.

\section{Conclusion}

We employed the structural parameters of metamaterials,
as magnetic oscillation strength, magnetic resonance frequency and magnetic damping constant,
to investigate their influence on the refractive index of metamaterials
and modes' behavior in metamaterial-based waveguides.
We also analyzed the robustness of metamaterials to the errors in the
metamaterial structural parameters.
Our survey shows that the magnetic oscillation strength
has the ability to increase or decrease the negative refractive index frequency region
and hybrid-mode region, whereas,
magnetic resonance frequency shifts this frequency region.
The magnetic damping affects value of refractive index as by increasing magnetic damping
the real part of refractive index gets more negative value.

Our investigations verify the effectiveness of metamaterials structural parameters on the double-negative refractive index frequency region of metamaterials and modes' behavior in metamaterial waveguides.
The results in this paper give an intuition to the choice of metamaterial unit-cell parameters for experimental construction of user intended metamaterial-based waveguides.

\section{Acknowledgement}

B.C.S.~acknowledges financial support from the NSERC, AITF, and China's
1000 Talent Plan and
N.S.~acknowledges financial support from NSERC. B.R.L.~acknowledges financial support from AITF.

%\section*{References}
\bibliography{mybibfile2}

\begin{thebibliography}{10}
\expandafter\ifx\csname url\endcsname\relax
  \def\url#1{\texttt{#1}}\fi
\expandafter\ifx\csname urlprefix\endcsname\relax\def\urlprefix{URL }\fi
\expandafter\ifx\csname href\endcsname\relax
  \def\href#1#2{#2} \def\path#1{#1}\fi

\bibitem{veselago2003electrodynamics}
V.~G. Veselago, Electrodynamics of materials with negative index of refraction,
  Phys. Usp. 46~(7) (2003) 764--768.

\bibitem{landy2008perfect}
N.~I. Landy, S.~Sajuyigbe, J.~J. Mock, D.~R. Smith, W.~J. Padilla, Perfect
  metamaterial absorber, Phys. Rev. Lett. 100~(20) (2008) 207402.

\bibitem{pendry2000negative}
J.~B. Pendry, Negative refraction makes a perfect lens, Phys. Rev. Lett.
  85~(18) (2000) 3966.

\bibitem{chen2012metamaterials}
T.~Chen, S.~Lin, H.~Sun, Metamaterials application in sensing, Sensors 12~(3)
  (2012) 2742--2765.

\bibitem{cai2007optical}
W.~Cai, U.~K. Chettiar, A.~V. Kildishev, V.~M. Shalaev, Optical cloaking with
  metamaterials, Nature Photon. 1~(4) (2007) 224--227.

\bibitem{pendry2006controlling}
J.~B. Pendry, D.~Schurig, D.~R. Smith, Controlling electromagnetic fields,
  Science 312~(5781) (2006) 1780--1782.

\bibitem{yeh2008essence}
C.~Yeh, F.~I. Shimabukuro, The Essence of Dielectric Waveguides, Springer, New
  York, 2008.

\bibitem{sang2015electromagnetic}
N.~Sang-Nourpour, B.~R. Lavoie, R.~Kheradmand, M.~Rezae, B.~C. Sanders,
  Electromagnetic-magnetoelectric duality for waveguides, arXiv:1510.06458.

\bibitem{tong2014advanced}
X.~C. Tong, Advanced Materials for Integrated Optical Waveguides, Springer, New
  York, 2014.

\bibitem{luo2002all}
C.~Luo, S.~G. Johnson, J.~D. Joannopoulos, J.~B. Pendry, All-angle negative
  refraction without negative effective index, Phys. Rev. B 65~(20) (2002)
  201104.

\bibitem{d2005te}
G.~D'Aguanno, N.~Mattiucci, M.~Scalora, M.~J. Bloemer, Te and tm guided modes
  in an air waveguide with negative-index-material cladding, Phys. Rev. E.
  71~(4) (2005) 046603.

\bibitem{wang2007nanoscale}
C.~J. Wang, L.~Y. Lin, Nanoscale waveguiding methods, Nanoscale Res. Lett.
  2~(5) (2007) 219--229.

\bibitem{lavoie2012low}
B.~R. Lavoie, P.~M. Leungand, B.~C. Sanders, Low-loss surface modes and lossy
  hybrid modes in metamaterial waveguides, Phot. Nano. Fund. Appl. 10~(4)
  (2012) 602--614.

\bibitem{shadrivov2012metamaterials}
I.~V. Shadrivov, P.~V. Kapitanova, S.~I. Maslovsk, Y.~S. Kivshar, Metamaterials
  controlled with light, Phys. Rev. Lett. 109~(8) (2012) 083902.

\bibitem{slobozhanyuk2014nonlinear}
A.~P. Slobozhanyuk, P.~V. Kapitanova, D.~S. Filonov, D.~A. Powell, I.~V.
  Shadrivov, M.~Lapine, P.~A. Belov, R.~C. McPhedran, Y.~S. Kivshar, Nonlinear
  interaction of meta-atoms through optical coupling, Appl. Phys. Lett. 104~(1)
  (2014) 014104.

\bibitem{rose2011overcoming}
A.~Rose, D.~R. Smith, Overcoming phase mismatch in nonlinear metamaterials,
  Opt. Mater. 1~(7) (2011) 1232--1243.

\bibitem{lapine2004three}
M.~Lapine, M.~Gorkunov, Three-wave coupling of microwaves in metamaterial with
  nonlinear resonant conductive elements, Phys Rev E Stat Nonlin Soft Matter
  Phys 70~(6) (2004) 066601.

\bibitem{powell2007self}
D.~A. Powell, I.~V. Shadrivov, Y.~S. Kivshar, M.~V. Gorkunov, Self-tuning
  mechanisms of nonlinear split-ring resonators, Appl. Phys. Lett. 91~(14)
  (2007) 144107.

\bibitem{lapine2009structural}
M.~Lapine, D.~Powell, M.~Gorkunov, I.~Shadrivov, R.~Marqu{\'e}s, Y.~S. Kivshar,
  Structural tunability in metamaterials, Appl. Phys. Lett. 95~(8) (2009)
  084105.

\bibitem{liu2012micromachined}
A.~Q. Liu, W.~M. Zhu, D.~P. Tsai, N.~I. Zheludev, Micromachined tunable
  metamaterials: a review, J. Opt. 14~(11) (2012) 114009.

\bibitem{rizza2015reconfigurable}
C.~Rizza, A.~Ciattoni, F.~D. Paulis, E.~Palange, A.~Orlandi, L.~Columbo,
  F.~Prati, Reconfigurable photoinduced metamaterials in the microwave regime,
  J. Phys. D: Appl. Phys. 48~(13) (2015) 135103.

\bibitem{penciu2010magnetic}
R.~S. Penciu, M.~Kafesaki, T.~Koschny, E.~N. Economou, C.~M. Soukoulis,
  Magnetic response of nanoscale left-handed metamaterials, Phys. Rev. B
  81~(23) (2010) 235111.

\bibitem{cai2010optical}
W.~Cai, V.~M. Shalaev, Optical Metamaterials, Springer, New York, 2009.

\bibitem{kamli2008coherent}
A.~Kamli, S.~A. Moiseev, B.~C. Sanders, Coherent control of low loss surface
  polaritons, Phys. Rev. Lett. 101~(26) (2008) 263601.

\bibitem{iizuka2002elements}
K.~Iizuka, Elements of Photonics, In Free Space and Special Media, Vol.~1, John
  Wiley \& Sons, New York, 2002.

\bibitem{ruppin2001surface}
R.~Ruppin, Surface polaritons of a left-handed material slab, J. Phys. Condens.
  Matter 13~(9) (2001) 1811.

\bibitem{Robustbook}
C.~Alippi, Intelligence for Embedded Systems, chapter 5, Robustness Analysis,
  Springer, New York, 2014.

\bibitem{Khurgin2015}
J.~B. Khurgin, How to deal with the loss in plasmonics and metamaterials, Nat.
  Nanotechnol. 10~(1) (2015) 2--6.

\end{thebibliography}
\end{document}